\documentclass[ag]{copernicus}

\usepackage{color}

\begin{document}

\title{{A note on the structure and kinematics of Harris current sheets}}

\author[1,2]{{R. A. Treumann}
}
\author[3]{{W. Baumjohann}}

\affil[1]{Department of Geophysics and Environmental Sciences, Munich University, Munich, Germany}
\affil[2]{International Space Science Institute, Bern, Switzerland}
\affil[3]{Space Research Institute, Austrian Academy of Sciences, Graz, Austria}

\runningtitle{Current sheet structure}

\runningauthor{R. A. Treumann and W. Baumjohann}

\correspondence{R. A.Treumann\\ (rudolf.treumann@geophysik.uni-muenchen.de)}

\received{ }
\revised{ }
\accepted{ }
\published{ }


\firstpage{1}

\maketitle

\begin{abstract}
The flat Harris current sheet model has become a powerful initial equilibrium in theory and simulation of magnetic reconnection in the magnetotail, at the magnetopause and also in multiple current sheet models of the heliosheath and astrophysics, where it is believed that such structures may become responsible for generation of turbulence and also high energy particles. Here we investigate the philosophy behind the Harris sheet in view of the physical conditions. The kinematic treatment in this note takes care of the different dynamics of ions and electrons.

\end{abstract}

\introduction

In the present communication a critical kinematic examination of the one-dimensional Harris current layer is undertaken from the point of view of its physical background. Such an investigation becomes necessary considering the many new observational and theoretical results of reconnection mostly achieved in the magnetospheric tail-current sheet, the majority of which do not come out as ingredients of the original Harris layer \citep{harris1962} or its refinements \citep[for a formal collection of distribution functions related to the Harris sheet see, e.g.,][]{balikhin2008} starting almost immediately \citep[with a paper by][]{bertotti1963}. The Harris sheet in either form is indeed a useful equilibrium model as long as overall pressure balance, quasi-neutrality and current sharing are realized. The latter implies that $u_i/T_i+u_e/T_e=0$ where the $u$s and $T$s are ion and electron drift velocities and temperatures, respectively. In addition, it does not distinguish between the different dynamics of the particle populations (ion, electrons) even not explicitly in the above cited work. under these assumptions the magnetic field assumes the simple $\tanh$-structure, and the current $\mathbf{j}(z)=\nabla\times\mathbf{B}(z)/\mu_0$ and density profiles $N(z)$ become identical:
\begin{equation}
B(z)=B_0\tanh\, (z/L), \qquad N=N_0\ \mathrm{sech}^2(z/L)
\end{equation}
where $B_0$ is the ambient magnetic field strength at $z\to\infty$, and $N_0$ the density maximum in the center of the current layer, and $L$ is the half-width of the current sheet which, here, is a free parameter applying to either narrow or thick current layers. Assuming relativistic conditions, i.e. relativistic current speeds, not relativistic temperatures, does not change this picture except for minor modifications \citep[see again, e.g.,][]{balikhin2008}. In retrospect, this has given justification for the use of the Harris model in simulations and also to construct multi-current layers adjacent to each other in order to investigate current layer interaction, believed to be the cause of plasma turbulence in multi-current layer carrying plasmas and of particle acceleration in interacting current sheets. The overall value of such investigations remains undisputed. 

The situation changes, if any of these assumptions is relaxed. From a physical point of view this has to be done unless the current sheet is fat, i.e. its half-width $L$ by far exceeds any of the relevant internal scales, in particular the ion and electron (indices $i,e$) gyroradii $\rho=\sqrt{2mT_{\perp}/e^2B^2}$ and the corresponding inertial scales $\lambda=\sqrt{m/\mu_0e^2N}$, with $(m,T,N)_{i,e}$ mass, temperature and density, respectively. This applies to dense current sheets embedded into moderately strong magnetic fields, $\mathbf{B}$, i.e. strong but distributed currents, $\mathbf{j}=e(N_i\mathbf{u}_i-N_e\mathbf{u}_e)=\nabla\times\mathbf{B}/\mu_0$, with $\mathbf{u}$ the bulk streaming velocity of the population. 

In collisionless plasmas like the tail of the magnetosphere one is, however, preferentially interested in thin current sheets the thickness $L\sim\lambda_i$   of which is of the order of the ion inertial length scale. Such current sheets, while being of vital interest in reconnection, have also been described by the Harris sheet model as initial condition. In the following we briefly discuss their physical properties in view of a distinction between the different dynamics of the current carrying ion and electron populations.

\section{Scalings}
When dealing with collisionless current sheets, it is convenient to define the  perpendicular electron and and ion plasma betas 
\begin{equation}
\beta_{(e,i)\perp}= 2\mu_0 N_{(e,i)}T_{(e,i)\perp}/B^2,  \qquad \beta_{i\perp}=\nu\theta\beta_{e\perp}
\end{equation}
The second equation expresses their relation, with  $\nu\equiv N_i/N_e$ the ion-to-electron density ratio, $\theta\equiv T_{i\perp}/T_{e\perp}$ the perpendicular ion-to-electron temperature ratio. With mass ratio $\mu=m_i/m_e$ we then have for the inertial lengths and gyroradii, respectively,
\begin{equation}
 \lambda_i=\sqrt{\mu\nu}\lambda_e, \qquad \rho_i=\sqrt{\mu\theta}\rho_e=\nu\sqrt{\mu\theta\beta_{e\perp}}\lambda_e
\end{equation}
Under stationary conditions it is reasonable (though not completely justified) to maintain the quasineutrality condition which reduces to $\nu=1$ yielding $\beta_{i\perp}/\beta_{e\perp}=\theta$ and
\begin{equation}
\frac{\lambda_i}{\lambda_e}=\sqrt{\mu},\ \ \frac{\rho_i}{\rho_e}=\sqrt{\mu\theta}, \ \ \frac{\rho_i}{\lambda_e}=\sqrt{\mu\theta\beta_{e\perp}}, \  \ \frac{\rho_e}{\lambda_e}=\sqrt{\beta_{e\perp}}
\end{equation}
For $\beta_{e\perp}=1$ one has $\rho_e=\lambda_e$ and, for $\theta=1$, also $\rho_i=\lambda_i$. 

The quasi-neutral pressure equilibrium of the Harris sheet model is equivalent to the assumption of a total $\beta$ 
\begin{equation}
\beta\equiv\ \beta_i+\beta_e\ =1
\end{equation}
where, in principle, $\beta=\beta_\perp+\beta_\|$. Thus the plasma is
everywhere throughout the sheet in local pressure balance; by accounting for the total pressures only it ignores the physically different effects along and transverse to the magnetic field (though in some treatments this is corrected by using temperature anisotropies $A=T_\perp/T_\|\neq 1$). This is not unreasonable though does not distinguish between the partial contributions of the two different populations. Such differences are vital, however, in the case when inertial effects dominate. The above expression, moreover, implicitly assumes that the current sheet is throughout magnetized which is true only under the additional condition that $\rho_e<\lambda_e$ everywhere, even in the center of the current sheet. This condition implies that $\beta_e<1$, in other words: no electron ``diffusion''region exists. Under quasi-neutrality this implies a condition on the perpendicular electron temperature
\begin{equation}\label{eq-cond}
T_{e\perp} \ , < \,  B^2/2\mu_0N \,Ê = \,{\textstyle\frac{1}{2}}\,m_iV_A^2
\end{equation}
The electron temperature (in energy units) is required to be less than the Alfv\'en energy $m_iV_A^2/2$ per ion.

Thinking of application to the magnetospheric tail current sheet, observations suggest that in the tail plasma sheet $\theta\approx 7$ \citep{baum1989,baum2011,kauf2005}, the cause of which is not well known yet. The current sheet electrons might be somewhat hotter indeed, possibly caused by internal processes like acceleration in the electron ``diffusion'' region, instabilities, or the necessity to balance the excess magnetic pressure at the boundary of the electron ``diffusion'' region (see below). In any case, it indicates that  electrons and ions are \emph{not} in local thermal equilibrium. The last equation for a magnetized Harris tail current sheet then implies that in the terrestrial magnetotail
\begin{equation}
T_i\sim 7\ \left({\textstyle\frac{1}{2}}m_iV_A^2\right) \qquad\mathrm{(magnetotail)}
\end{equation}
For densities $1<N<10$ m$^{-3}$ and fields $0.1 < B <10$ nT we have $1 <T_i <  100$ keV, an interval large enough to contain all observed values and thus being undecidable. The undisturbed current sheet in the magnetospheric tail contains, moreover, a small positive normal residual dipole field component $+B_z\ne0$ not resembling the ideal Harris sheet used in most simulations. 

At the magnetopause at the contrary, nominally $N\sim 3\times 10^7$ m$^{-3}$, $B\sim 30 $ nT, and presumably $\theta\sim 1$, yielding $T_i\sim 150$ eV, showing that the magnetopause, except for its magnetic asymmetry, probably lacks any non-magnetic electron inertial ``diffusion''  region. It more resembles a completely magnetized Harris sheet even on its magnetosheath side. 

Returning to the magnetotail, for a proton-electron plasma, $\sqrt{\mu}=43$ and thus $\lambda_i\approx 43\,\lambda_e,\ \rho_i\approx 110\,\rho_e$. In this particular case the ion gyroradius-to-inertial scale ratio exceeds the same ratio for electrons, i.e. ions become dominated by inertial effects on scales $\lambda_i<\rho_i$; similarly electrons are nonmagnetic on scales $\lambda_e<\rho_e$. In a bulk plasma this is of little interest. However, in a current sheet whose thickness matches one of these scales the distinction between the different plasma behavior becomes vital, when the scales are measured from the center of the current layer. This is all well known \citep[it has first been discussed by][in connection to its effects in magnetic reconnection]{sonnerup1979}, and we can be very brief in the following when referring to the consequences. 

\section{Ion ``diffusion" region}
Here we have $\rho_i>\lambda_i$ as measured from the center of the current layer at $z=0$, and the ions decouple from the magnetic field and start behaving non-magnetic while the electrons remain to be magnetic outside a distance $z=\lambda_e$. Once a cross-field electric field $\mathbf{E}=E_y{\hat y}$ exists in this region, then the electrons experience it continuing their $E\times B$-drift toward the center of the current sheet. Since the ions do not follow them anymore under quasi-neutral conditions, the electron drift motion corresponds to a normal Hall current 
\begin{equation}
J_{Hz}=eN(z)E_y/B_x(z)
\end{equation}
the magnetic field of which adds perpendicularly to the ambient field pointing into the direction of the ambient convection electric field. In the plane undisturbed Harris sheet this current is directed away from the current plane along $\pm z$ with the upper sign applying above, the lower below the sheet. This is the first and now well-known effect which \citet{sonnerup1979} realized and which has subsequently been used and referred to in observations \citep[][and others]{fujimoto1997,nagai2001,oieroset2001} and simulations. 
The current strength is 
\begin{equation}
J_{Hz}=\frac{eN_0}{B_0}\left[\sinh\left(\frac{2z}{L}\right)\right]^{-1}\!\!\!\!\!\!\!\!\!\!\!\!\!\!\!, \qquad \lambda_e<|z|<\lambda_i.
\end{equation}
Putting $L\sim\lambda_i$ one realizes that the Hall current is strongest in the \emph{inner part} of the ion ``diffusion'' region, a dependence on $z$ caused by the increase of the magnetic field with distance from the sheet. These Hall currents close presumably along the magnetic field in $\pm x$-direction. In a plane Harris sheet, no closure current direction is preferred. 

The electrons, being magnetized, also feel the cross-tail electric field $E_y$ in another way: this field points along the Hall magnetic field. As a consequence, in the regions of the Hall magnetic field the electrons become accelerated across the tail against the flow of the ions to energies
\begin{equation}
\epsilon_e\sim eE_y\ell
\end{equation}
where $\ell$ is the length over which the electric field is along the Hall field. This length may be assumed to be of the order of the ion inertial length, $\ell\gtrsim\lambda_i$. The energy of flow, for $E_y\sim 1$ mV/m, and $N_0 \sim 10^5$ m$^{-3}$, can reach values of many keV into backward direction $-y$, corresponding to substantial flow velocities of $\gtrsim 10^4$ km/s and causing the electrons to develop an energetic tail as well as  to amplify the cross tail current locally in the region of the Hall field, independent of the sign of the Hall-field direction. This causes some striation of the cross tail current. Since in ongoing reconnection the Hall field is strongest near the separatrices it leads to partial bifurcation of the cross tail current in the ion ``diffusion'' region, a simple effect of the presence of Hall currents and the separation between ion and electron inertia.

Finally, the separation  of  ion and electron motion in the ion ``diffusion'' region causes electron drift currents. These are also most pronounced in the inner part of the ion ``diffusion'' region where the magnetic field gradient is strongest. These currents are carried solely by electrons and, similar to the ring current, are not compensated by pressure effects or by the ion component. Their magnitude is
\begin{equation}
\mathbf{J}_{\nabla B}= {\hat y}\, N(z)T_{e\perp}\big[ \nabla_z B(z)\big]/B^2(z)
\end{equation}
In the Harris sheet one thus has another small positive drift-current contribution to the tail current 
\begin{equation}
J_{\nabla B, y}(z) \sim \frac{N_0T_{e\perp}}{2B_0L}\left[\sinh\left(\frac{2z}{L}\right)\right]^{-2}\!\!\!\!\!\!\!\!\!\!\!\!,\quad \lambda_e<|z|\lesssim\lambda_i
\end{equation}
in the inner part of the ion ``diffusion'' region. If we put $L\sim\lambda_i$, it becomes clear that in a plane Harris current sheet this drift current maximizes near $|z|\gtrsim \lambda_e$, where it contributes to amplification of the tail current. One may note that the original Harris sheet does not refer to any of these currents and effects. 

In addition, under reconnection the non-magnetic ions also feel the cross-tail convection electric field and over the ion-inertial lengths become accelerated across the tail to an energy
\begin{equation}
\epsilon_i\simeq eE_y\lambda_i
\end{equation}
In the magnetospheric tail this number can reach values of the order up to $\epsilon_i\lesssim 1$ MeV.

\section{Electron ``diffusion" region}
The most interesting region in an ideal nonmagnetic Harris sheet, where the condition Eq. (\ref{eq-cond}) is violated, is the electron ``diffusion'' region. It is located at distance $|z|<\lambda_e$ from the center of the tail current. Here neither the ions nor the electrons are magnetic. The electrons have decoupled from the magnetic field as well. As a bulk component they feel the cross-tail electric field, become accelerated across the tail and contribute massively to the tail current. However, what is more interesting, is the fact that the electron ``diffusion'' region represents an \emph{ideally conducting} layer of thickness $\gtrsim2\lambda_e$. No magnetic field can penetrate here except over the electron skin-depth $\lambda_e\lesssim d_e$ with magnetic  field varying according to
\begin{equation}
B_x(z)=\pm B_x(d_e)\exp\left(-|z-d_e|/\lambda_e\right), \qquad |z|<d_e
\end{equation}
The magnetic field inside the electron ``diffusion'' region decays \emph{exponentially} toward the center of the current sheet. Instead, the Harris sheet model suggests a linear decay, which is due to the complete neglect of the inertial effects on the electrons. In addition, in a \emph{non-driven} Harris sheet, pressure balance at the boundary of the electron ``diffusion'' region requires for equilibrium that
\begin{equation}
N_e(d_e) = B_x^2(d_e)/2\mu_0T_e \qquad |z|<d_e
\end{equation}
assuming constant electron temperature. This yields in a Harris sheet a small excess in electron density
\begin{equation}
N_e\approx B_0^2\tanh^2(d_e/L)/2\mu_0T_e \sim N_0 \tanh^2(d_e/L)
\end{equation}
where use has been made of the Harris isothermal assumption. With $L\sim\lambda_i, d_e\sim\lambda_e$ this becomes 
\begin{equation}
N_e \sim N_0\tanh^2 (\mu^{-1})\approx \mu^{-2}N_0\ \lesssim 10^{-3}N_0, \  \ |z|<\lambda_e
\end{equation}
an effect which is not remarkable in the density. In spite of a density enhancement due to compression the excess magnetic pressure can also be balanced by an enhancement in electron temperature, an indication of which might be seen in the above mentioned \citep{baum2011} slightly lower ion-to-electron temperature ratio than in the surrounding tail plasma sheet. Though the density effect is negligible, the magnetic effect is not, because due to exponential exclusion of the magnetic field from the electron ``diffusion'' region the center of current sheets is practically free of magnetic fields, which has profound consequences on mechanisms like reconnection which require that the anti-parallel magnetic fields separated by the current layer must come into mutual contact in order to reconnect.  

\section{Driven electron `diffusion'' layer}
More interesting is the driven case which happens when the current layer is embedded into plasma inflow as, for instance, in the magnetotail. There the flow $\mp v_z=E_y/(\pm B_x)$ is caused by the presence of the electric convection field $+E_y$, continuously transporting plasma and field into the plasma sheet and current layer from where the plasma must flow out, under stationary conditions electrons being squeezed (by the $E_y$-field) into $- y$-direction (carrying current to infinity). Any continuous inflow will necessarily cause a pile-up of the field at the edge of the electron ``diffusion'' region $d_e$. This process exerts growing pressure compressing the ``diffusion'' region until it ultimately disappears. When this happens, the current layer becomes completely magnetic, with electron orbits turning into Speiser orbits. One can estimate this process in the following way:  The evolution of the magnetic field follows from the collisionless (i.e., diffusionless) induction equation $\partial_t \mathbf{B}(t)=\nabla\times\mathbf{v}\times\mathbf{B}$, for $\mathbf{B}=B_x\hat x$ yielding dimensionally
\begin{equation}\label{eq-ind}
\frac{\partial B_x}{\partial t}= v\frac{\partial B_x}{\partial z} \sim \frac{v}{\lambda_e}B_x
\end{equation}
when accounting for the variation of the magnetic field across the skin depth only, with time $t$ (or $d_e-|z|$) measured from the boundary of the electron ``diffusion'' region, $|z|=d_e$. Pressure balance requires (for constant $T$) that
\begin{equation}
\frac{N(d_e-|z|)}{N_0}= \frac{B_x^2(d_e-|z|)}{B_0^2}, \qquad |z|<d_e
\end{equation}
at each point $|z|<d_e$ during the evolution of the $B$ field, density $N$ and width of the electron ``diffusion'' region. Measuring $\zeta=|z|/\lambda_e, \alpha=d_e/\lambda_e$ in skin depths we have for the magnetic field penetrating the electron ``diffusion'' region
\begin{equation}
B_x(\alpha-\zeta)=B(\alpha)\exp\left[(\alpha-\zeta)\sqrt{N(\alpha-\zeta)}\right]
\end{equation}
with $B(\alpha)=B_0\tanh(\alpha\lambda_e/L)$. Replacing in Eq. (\ref{eq-ind}) $\lambda_e=\lambda_e(\alpha)\sqrt{N(\alpha)/N(\alpha-\zeta)}$ and eliminating the density yields 
\begin{equation}
B_x^{-2}\frac{\partial B_x}{\partial t} =\frac{v}{B(\alpha)\lambda_e(\alpha)}
\end{equation}
The expression on the right does not explicitly depend on time. So we can integrate it  -- under the condition that at time $t=0$ the magnetic field is $B_x(t=0)=B(\alpha)$ -- to obtain
\begin{equation}
B_x(t)\simeq \frac{B(\alpha)}{1-t/t_0}, \qquad t<\lambda_e(\alpha)/v \equiv t_0(\alpha)
\end{equation}
for $B_x(t)$, and from pressure balance 
\begin{equation}
N(t)=\frac{N(\alpha)}{[1-t/t_0]^2}
\end{equation}
Obviously field and density in the electron ``diffusion'' region increase with time. Hence, the electron skin depth evolves according to 
\begin{equation}
\lambda_e(t)=\lambda_e(\alpha)(1-t/t_0)
\end{equation}
vanishing at $t=t_0$. At this time the electron ``diffusion'' region disappears, and the current layer becomes completely magnetic. Using this expression in the magnetic field we obtain the evolution of the magnetic field as
\begin{equation}
B_x(t)=B(\alpha)\exp \left[\frac{\alpha-\zeta(\alpha)}{1-t/t_0}\right], \qquad \zeta < \alpha
\end{equation}
This (approximate) expression shows the gradual increase with time of the magnetic field in the electron ``diffusion'' region during continuous inflow of plasma and magnetic field at constant speed $v$ and local pressure balance.  Apparently the magnetic field  explodes for $t=t_0$. This is, however, not the case because near $t\sim t_0$ one has $\lambda(t_0)\to0$ and, hence, $\zeta\sim\alpha$ and $B_x(t_0)\sim B(\alpha)$. At this time, the magnetic field has filled the current layer at about its value outside the original electron ``diffusion'' layer, and the ``diffusion'' region has  completely disappeared. The oppositely directed fields of non-zero strengths $B_x=\pm B(\alpha)$ contact each other, and the current necessarily becomes unstable setting on reconnection in some way.

\section{Discussion}
The real physical structure of the Harris sheet is more complicated than the original and more sophisticated models suggest. These neglect the competition between the inertial and magnetic effects in the current sheet. Such effects do not occur in the bulk plasma; however, in the current sheet the center of the sheet is a singular layer which makes the difference between inertia and magnetism being felt by the plasma as had first been realized by \citet{sonnerup1979}. Accounting for these differences becomes necessary when dealing with reconnection. The first and important effect, pointed out by \citet{sonnerup1979} and later being observed \citep[first by][followed by others]{fujimoto1997,nagai2001,oieroset2001,naka2006,runov2003,runov2003a} and confirmed by various PIC simulations \citep[][and others]{zeiler2002,drake2003,ricci2004} is the Hall effect, the importance of which lies in the decoupling of the electron and ion fluid and in the generation of field-aligned currents which, in the case of the magnetosphere, close in the ionosphere and lead to the various effects observed in aurorae during substorms and storms. Presumably Hall currents have little effect on the reconnection process itself but contribute to the currents noted above, guide fields, and particle acceleration. 

The other still badly understood effects relate to the electron inertial region \citep{scudder2008}. As we have argued here these effects keep the current sheet centre in the magnetically undriven case free of  magnetic fields. Thus, in order to obtain sufficiently fast reconnection, other processes must be invoked in order to transport magnetic fields into the center. Such processes can either be quantum mechanical \citep{treu2013} though are probably difficult to confirm experimentally, or require forcing or driving of reconnection. Such kind of forcing is naturally provided either by the presence of a sufficiently strong guide field \citep{baum2010,treu2010} and/or electric convection fields $E_y$ which are responsible for inward transport of magnetic field into the current layer from both sides of the antiparallel magnetic fields. This inward transport gradually piles up magnetic field lines and pressure at the boundary of the electron inertial ``diffusion" region thereby secondarily magnetizing the electrons here and compressing the electron inertial region until it shrinks to an extension less than the electron gyroradius. This causes the inertial region to disappear and restore, under forcing conditions, a completely magnetized current layer resembling the originally assumed Harris sheet, this time, however, under forcing conditions and bringing the oppositely directed magnetic fields into mutual contact. We have illustrated this process here with a rough estimate showing the gradual increase of the magnetic field towards the center of the current sheet at the same time when the electron inertial region shrinks in extension. Reconnection will then spontaneously set on. Such processes are also expected both at the magnetopause where the electron ``diffusion'' layer is anyway either absent or very thin, and in the tail current sheet under the assumed forcing.

\begin{acknowledgements}
This research was part of an occasional Visiting Scientist Programme in 2006/2007 at ISSI, Bern.
\end{acknowledgements}


\begin{thebibliography}{ }

\bibitem[Artemyev et al.(2011)]{baum2011} Artemyev, A. V., Baumjohann, W., Petrukhovich, A. A., Nakamura, R., Dandouras, I.  \& Fazakerley, A.: Proton/electron temperature ratio in the magnetotail, Ann. Geophys. 29, 2253-2257, doi: 10.5194/angeo-29-2253-2011, 2011. 

 
 \bibitem[Balikhin and Gedalin(2008)]{balikhin2008} Balikhin, M. A. \& Gedalin, M.: Generalization of the Harris current sheet model for non-relativistic, relativistic and pair plasmas, J. Plasma Phys. 74, 749-763, doi: 10.1017/S002237780800723X, 2008.

\bibitem[Baumjohann et al.(1989)]{baum1989} Baumjohann, W., Paschmann, G. \& Cattell, C. A.: Average plasma properties in the central plasma sheet, J. Geophys. Res. 94, 6597-6606, doi: 10.1029/JA094iA06p06597, 1989. 


 
\bibitem[Baumjohann and Treumann(2012)]{treumann1996} Baumjohann, W. \& Treumann, R. A.: Basic Space Plasma Physics, Revised edition, Imperial College Press, London 2012. 

\bibitem[Baumjohann et al.(2010)]{baum2010} Baumjohann, W., Nakamura, R. \& Treumann, R. A.: Magnetic guide field generation in collisionless current sheets, Ann. Geophys. 28, 789-793, doi: 10.5194/angeo-28-789-2010, 2010. 


\bibitem[Bertotti(1963)]{bertotti1963} Bertotti, B.: Fine structure in current sheets, Ann. Phys. 25, 271-289, doi: 10.1016/0003-4916(63)90014-9, 1963.



\bibitem[Drake et al.(2003)]{drake2003} Drake, J.~F., Swisdak, M., Cattell, C., Shay, M.~A., Rogers, B.~N. \& Zeiler, A.: Formation of electron holes and particle energization during magnetic reconnection,  Science 299, 873-877, doi: 10.1126/science.1080333, 2003.


\bibitem[Fujimoto et al.(1997)]{fujimoto1997}
Fujimoto, M., Nakamura, M. S., Shinohara, I., Nagai, T., Mukai, T., Saito, Y., Yamamoto, T. \& Kokubun, S.: Observations of earthward stre\-aming electrons at the trailing boundary of a plasmoid, Geophys. Res. Lett. 24, 2893-2896, doi: 10.1029/97GL02821, 1997.


\bibitem[Harris(1962)]{harris1962} Harris, E. G.: On a plasma sheath separating regions of oppositely directed magnetic fields, Nuovo Cim. 23, 115-121, 1962.

\bibitem[Kaufmann et al.(2005)]{kauf2005} Kaufmann, R. L., Paterson, W. R. \& Frank, L. A.: Relationships between the ion flow speed, magnetic flux transport rate, and other plasma sheet parameters, J. Geophys. Res. 110, A09216, doi: 10.1029/2005JA011068, 2005. 



\bibitem[Nagai et al.(2001)]{nagai2001}   Nagai, T., Shinohara, I., Fujimoto, M., Hoshino, M., Saito, Y., Machida, S. \& Mukai, T.:   Geotail observations of the Hall current system: Evidence of magnetic reconnection in the magnetotail,   J. Geophys. Res.   106,  25929-25950, doi: 10.1029/2001JA900038,   2001.

\bibitem[Nakamura et al.(2006)]{naka2006} Nakamura, R., Baumjohann, W., Asano, Y., Runov, A., Balogh, A., Owen, C. J., Fazakerley, A. N., Fujimoto, M., Klecker, B. \& R\`eme, H.: Dynamics of thin current sheets associated with magnetotail reconnection,  J.  Geophys. Res. 111, A11206, doi:10.1029/2006JA011706, 2006.



\bibitem[{\O}ieroset et al.(2001)]{oieroset2001}  { {\O}ieroset, M., Phan, T. D., Fujimoto, M., Lin, R. P. \& Lepping, R. P.}:   In situ detection of collisionless reconnection in the Earth's magnetotail,  {Nature}{ 412},   {414-417, doi: 10.1038/35086520},  {2001}.


\bibitem[Ricci et al.(2004)]{ricci2004} {Ricci, P., Brackbill, J. U., Daughton, W. \& Lapenta, G.}: Collisionless magnetic reconnection in the presence of a guide field, {Phys. Plasmas} {11}, {4102-4114, doi: 10.1063/1.1768552}, {2004}.



\bibitem[Runov et al.(2003a)]{runov2003} Runov, A., Nakamura, R., Baumjohann, W., Zhang, T.~L., Volwerk, M., Eichelberger, H.-U. 
\& Balogh, A.: Cluster observation of a bifurcated current sheet, Geophys. Res. Lett. 30, 1036,  doi:10.1029/2002GL016136, 2003a. 

\bibitem[Runov et al.(2003b)]{runov2003a} Runov, A., Nakamura, R., Baumjohann, W., Treumann, R. A., Zhang, T.~L., Volwerk, M., V\"or\"os, Z., Balogh, A., Glassmeier, K.-H., Klecker, B., R\`eme, H. \& Kistler, L.: Current sheet sstructure near magnetic X-line observed by Cluster, Geophys. Res. Lett. 30, 1579,  doi:10.1029/2002GL016730, 2003b. 

\bibitem[Scudder et al.(2008)]{scudder2008} {Scudder, J. D., Holdaway, R. D., Glassberg, R., and Rodriguez, S. L.}:  Electron diffusion region and thermal demagnetization, {J. Geophys. Res.} {113}, {A10208, doi: 10.1029/2008JA013361}, {2008}.



\bibitem[Sonnerup(1979)]{sonnerup1979} Sonnerup, B. U. \"O.: Magnetic field reconnection, in: Solar system plasma physics, Vol. III.,  pp. 45-108, edited by: Lanzerotti, L. T., Kennel, C. F., and Parker, E. N., North-Holland Publ. Comp., Amsterdam-New York, 1979.



\bibitem[Treumann et al.(2010)]{treu2010} Treumann, R.~A.,  Nakamura, R. \& Baumjohann, W.: Collisionless reconnection: mechanism of self-ignition in thin plane homogeneous current sheets, Ann. Geophys. 28, 1935-1943, doi:  10.5194/angeo-28-1935-2010, 2010.

\bibitem[Treumann et al.(2012)]{treu2013} Treumann, R.~A.,  Baumjohann, W. \& Gonzalez, W. D.: Collisionless reconnection: magnetic field line interaction, Ann. Geophys. 30, 1515-1528, doi:  10.5194/angeo-30-1515-2012, 2012.


\bibitem[Zeiler et al.(2002)]{zeiler2002} Zeiler, A., Biskamp, D., Drake, J. F., Rogers, B. N., Shay, M. A. \& Scholer, M.: Three-dimensional particle simulations of collisionless magnetic reconnection, J. Geophys. Res. 107, A1230, doi: 10.100729/2001JA000287, 2002. 



\end{thebibliography}
\end{document}